\documentclass[12pt]{article}
\input{epsf}
\usepackage{times}
\author{Morteza Mohseni\thanks{email: m-mohseni@pnu.ac.ir}\\
\small Physics Department, Payame Noor University, 19395-4697
Tehran 19569, Iran\\
\small and\\
\small Institute for Studies in Theoretical Physics and
Mathematics (IPM),\\\small 19395-5531 Tehran, Iran}
\date{}

\begin{document}
\title{Spinning particles in gravitational wave spacetime}
\maketitle \abstract{The dynamics of pseudo-classical spinning
particles in spacetime of gravitational plane waves of general
polarization and harmonic profile is studied. The resulting
equations of motion are solved exactly and the results are
compared with those of the other approaches. The relative
accelerations of nearby particles is also calculated.} \vskip
0.5cm \noindent PACS numbers: 04.20.-q, 04.25.-g, 04.30.-Nk

\newpage
\section{Introduction}
The dynamics of particles in curved spacetime has been an important
part of the theories of gravitation. The dynamics of test particles, i.e.,
particles without any internal structure was studied in the early
stages of the development of general relativity. Recently, there has been a growing
interest in the problem of motion of spinning particles (some references may
be found in \cite{1}).

Rietdijk and van Holten \cite{3} has obtained a general set of
equations of the symmetries of the spinning particles in an
arbitrary curved spacetime. Similar equations has been obtained in
\cite{bar} from a more general action. These equations constitutes
a Grassmann valued extension of the Killing equations for the
symmetries of the spacetime manifold. These ideas was further
investigated in \cite{4} and was applied to the case of
Schwarzschild spacetime in \cite{5} and to Reissner-Nordstrom
spacetime in \cite{6}. The canonical structure of the theory  and
the equivalent Dirac formulation of the quantum theory has been
derived in \cite{wci}. Equations for the world-line deviations of
spinning particles was obtained in \cite{nik}.

The aim of the present paper is to study the motion of spinning
particles in the spacetime of an exact plane gravitational wave
following the ideas presented in the references mentioned above.
This problem has been studied previously in the framework of
Papapetrou-Dixon description of spinning particles in \cite{1} and
\cite{ma,7,kes}. In \cite{1} a linearized version of
Papapetrou-Dixon equations was applied to the case of a weak plane
gravitational wave of arbitrary polarization and harmonic profile.
In \cite{7} a linearly polarized wave of arbitrary profile (and in
particular, a square gravitational wave pulse) was considered and
a set of exact solutions was presented in some rather special
circumstances. The general solution of those equations could only
be found numerically, of course.

There is a simpler version of Papapetrou-Dixon equations. This is
basically obtained by replacing the usual supplementary condition
$p\cdot s=0$ by the so called Pirani supplementary condition
$v\cdot s=0$. A similar calculation has been performed along the
lines of \cite{1}, but using Pirani condition instead \cite{tul}.
It led to conclusions very close to those of \cite{1}.

In this paper we find the general solution for motion of spinning
particles in the spacetime of an exact plane gravitational wave.
We use only two of the generic constants of motion, namely the
Hamiltonian and the supercharge. The results may be considered as
a generalized alternative to the results of \cite{1} and \cite{7}.
The solutions we present here, give more physical insight into the
behaviour of spinning particles and may be of interest in the
important problem of the detection of gravitational waves. In the
subsequent sections we first give brief reviews of spinning
particles and gravitational waves and then solve the equations of
motion exactly, next we study the world-line deviations of
spinning particles and finally present our physical conclusions.
An appendix is devoted to the small-$h$ limit of the particle's
trajectory and the evolution of its spin.
\section{Spinning particles}
The equations of motion of a classical fermion may be obtained
from the action \cite{wci,3}
\begin{equation}
S=m\int d\tau(\frac{1}{2}g_{\mu\nu}(x){\dot x}^\mu{\dot
x}^{\nu}+\frac{i}{2m}
g_{\mu\nu}(x)\psi^\mu\frac{D\psi^\nu}{D\tau}),\label{eq0}
\end{equation}
$x^{\mu}$ being commuting and
$\psi^\mu$ anticommuting variables with $\mu,\nu,\cdots$ running over the
dimensions of the spacetime. The above action will lead to the following
equations of motion
\begin{eqnarray}
\frac{D^2x^\mu}{D\tau^2}&=&\frac{i}{2m}{R^\mu}_{\nu\lambda\rho}
{\dot x}^{\nu}\psi^\lambda\psi^\rho, \label{eq1}\\
\frac{D\psi^\mu}{D\tau}&=&0, \label{eq2}
\end{eqnarray}
where $\frac{D^2x^\mu}{D\tau^2}:={\ddot x}^\mu
+{\Gamma^\mu}_{\lambda\nu} {\dot x}^\lambda{\dot x}^\nu$,
$\frac{D\psi^\mu}{D\tau}:={\dot\psi}^\mu+{\dot x}^
{\lambda}{\Gamma^\mu}_{\lambda\nu}\psi^\nu$, a dot denotes
$\frac{d}{d\tau}$, and
${R^\nu}_{\kappa\lambda\mu}:=\partial_\lambda\Gamma^\nu_
{\kappa\mu}-\partial_\mu\Gamma^\nu_{\kappa\lambda}+\Gamma^\nu_{\lambda\rho}
\Gamma^\rho_{\kappa\mu}-\Gamma^\nu_{\mu\rho}\Gamma^\rho_{\kappa\lambda}$
is the Riemann curvature tensor.

There are four types of generic (i.e. existing in any spacetime) conserved
quantities connected with the action (\ref{eq0}). These are the Hamiltonian
$H$ ,the supercharge $Q$, the dual supercharge $Q^\star$, and the chiral
charge $\Gamma_\star$, given respectively by

\begin{eqnarray}
H&=&\frac{1}{2m}g^{\mu\nu}(x)p_\mu p_\nu=\frac{m}{2}g^{\mu\nu}(x){\dot x}_\mu
{\dot x}_\nu,\label{eq4}\\
Q&=&p_\mu\psi^\mu,\label{eq5}\\
Q^\star&=&\frac{1}{3!}\sqrt{-g}\epsilon_{\mu\nu\kappa\lambda}p^\mu\psi^\nu\psi^
\kappa\psi^\lambda,\label{eq6}\\
\Gamma_\star&=&\frac{1}{4!}\sqrt{-g}\epsilon_{\mu\nu\kappa\lambda}\psi^\mu
\psi^\nu\psi^\kappa\psi^\lambda,\label{eq7}
\end{eqnarray}
The relativistic spin of the particle is described by the antisymmetric tensor
\begin{equation}
s^{\mu\nu}=-i\psi^\mu\psi^\nu,\label{eq8}
\end{equation}
whose spacelike components $s^{ij}$ represent the particle's magnetic dipole
moment and the timelike components $s^{0i}$ correspond to the electric dipole
moment. For free Dirac particles we require the dipole moment to vanish in the
rest frame. Hence we set
\begin{equation}
g_{\mu\lambda}(x){\dot x}^{\lambda}s^{\mu\nu}=0, \label{eq9}
\end{equation}
or equivalently
\begin{equation}
g_{\mu\nu}(x){\dot x}^\mu\psi^\nu=0,\label{eq10}
\end{equation}
That is $Q=0$. The conservation of $Q$ guarantees that this
condition can be satisfied at all times irrespective of the
presence of external fields.

We may recast Eqs. (\ref{eq1}) and (\ref{eq2}) in terms of $s^{\mu\nu}$ as follows
\begin{eqnarray}
\frac{D^2x^\mu}{D\tau^2}&=&-\frac{1}{2m}R^{\mu}_{\hspace{2mm}\nu\lambda\rho}
{\dot x}^{\nu}s^{\lambda\rho}, \label{eq11}\\
\frac{Ds^{\mu\nu}}{D\tau}&=&0. \label{eq12}
\end{eqnarray}
where $\frac{Ds^{\mu\nu}}{D\tau}:={\dot s}^{\mu\nu}+\Gamma^\mu_{\kappa\lambda}
{\dot x}^\kappa s^{\lambda\nu}+\Gamma^\nu_{\kappa\lambda}{\dot x}^\kappa
 s^{\mu\lambda}.$
The particle's spin may also be represented via a spin four-vector given by
\begin{equation}
s^\mu:=\frac{1}{2\sqrt{-g}}\epsilon^{\mu\nu\kappa\lambda}{\dot x}_{\nu}
s_{\kappa\lambda}.\label{eq13}
\end{equation}
In general relativity, only relative acceleration has an observer-
independent meaning. The relative acceleration of two nearby
particles can be determined by measuring the rate at which the
particles world-lines deviate from each other. The world-line
deviation equations are given by
\begin{eqnarray}
\frac{D^2n^\mu}{D\tau^2}&=&-{R^\mu}_{\kappa\lambda\nu}{\dot
x}^\kappa {\dot x}^\nu n^\lambda-\frac{1}{2m}s^{ab}{R^\mu}_{\nu
ab}\frac{Dn^\nu}{D\tau}\nonumber\\&-&\frac{1}{2m}(s^{ab}\nabla_\lambda
{R^\mu}_{\nu ab}{\dot x}^\nu n^\lambda+J^{ab}{R^\mu}_{\nu ab}{\dot
x}^\nu),\label{eqd1}\\
\frac{DJ^{\mu\nu}}{D\tau}&=&s^{\kappa[\mu}{R^{\nu]}}_{\kappa\alpha\beta}n^\alpha{\dot
x}^\beta ,\label{eqd2}
\end{eqnarray}
where $\nabla_\lambda
{R^\mu}_{\nu\kappa\rho}=\partial_\lambda{R^\mu}_{\nu\kappa\rho}
+{\Gamma^\mu}_{\lambda\beta}{R^\beta}_{\nu\kappa\rho}
-{\Gamma^\beta}_{\nu\lambda}{R^\mu}_{\beta\kappa\rho}
-{\Gamma^\beta}_{\kappa\lambda}{R^\mu}_{\nu\beta\rho}
-{\Gamma^\beta}_{\rho\lambda}{R^\mu}_{\nu\kappa\beta}$ and
$n^\mu,J^{ab}$ are defined for a one-parameter congruence of
solutions $(x^\mu(\tau;\lambda),\psi(\tau;\lambda))$ to
(\ref{eq1}) and (\ref{eq2}) as follows
\begin{eqnarray*}
n^\mu=\frac{\partial x^\mu}{\partial\lambda},\hspace{1cm}
J^{ab}=\frac{Ds^{ab}}{D\lambda}.
\end{eqnarray*}
Equations (\ref{eqd1})-(\ref{eqd2}) are generalization of the
well-known equation of geodesic deviation. The first term in the
r.h.s of (\ref{eqd1}) represents the usual effect of the spacetime
curvature on the deviations of world-lines, and the other terms
are due to the spin-curvature coupling. These equations may be
derived by a procedure  similar to the derivation of the equation
of geodesic deviation, or by alternative methods \cite{nik}.

In the next section we seek solutions to Eqs. (\ref{eq9}),
(\ref{eq11}), (\ref{eq12}) and(\ref{eqd1}),(\ref{eqd2}) in a
gravitational wave spacetime.

\section{Plane gravitational  waves}
The most general plane-fronted parallel-rays gravitational wave is given by
the metric \cite{8}
\begin{equation}
ds^2=-dudv+F(u,x^a)du^2+dx^adx^a,\label{eq14}
\end{equation}
where $(u,v)$ are null coordinates, $x^a$ are spacelike transverse
coordinates, and $F(u,x^a)$ is a solution of the two-dimensional
Laplace equation with respect to $x^a$. Plane gravitational waves
are the particular case where $F$ is quadratic in $x^a$. Here we
consider a wave given by the metric
\begin{equation}
ds^2=-dudv-K(u,x,y)du^2+dx^2+dy^2,\label{eq15}
\end{equation}
where $K(u,x,y)=f_+(u)(x^2-y^2)+2f_\times(u) xy$. The arbitrary
functions $f_+$ and $f_\times$ corresponding to two independent
polarisations of the wave. We consider a wave of harmonic profile
(suitably chosen to reduce to the metric given in \cite{1} in the
weak field limit, in the so called group coordinates)
\begin{eqnarray}
f_+(u)&=&h\omega^2\sin(\omega u),\label{eq16a}\\
f_\times(u) &=& h\omega^2\cos(\omega u),\label{eq16b}
\end{eqnarray}
$h$ being the dimensionless wave amplitude.
\section{The particle's trajectory and spin}
In this section we seek solutions to Eqs. (\ref{eq9}) and (\ref{eq11})-
(\ref{eq13}) together with (\ref{eq15}) and suitable initial conditions. $\mu
=1,2,3,4$ correspond to $u,v,x,y$ respectively. From Eq. (\ref{eq12}) we have
\begin{eqnarray}
\frac{ds^{13}}{d\tau}&=&0,\label{eq17}\\
\frac{ds^{14}}{d\tau}&=&0,\label{eq18}\\
\frac{ds^{34}}{d\tau}&=&-\frac{1}{2}K_x{\dot u}s^{14}+\frac{1}{2}K_y{\dot u}
s^{13}.\label{eq19}
\end{eqnarray}
The three remaining $s^{\mu\nu}$ may be found from Eq. (\ref{eq9}) with $\nu=1,3,4$
\begin{eqnarray}
{\dot u}s^{12}&=&2{\dot x}s^{13}+2{\dot y}s^{14},\label{eq20}\\
{\dot u}s^{23}&=&-({\dot v}+2K{\dot u})s^{13}-2{\dot y}s^{34},\label{eq21}\\
{\dot u}s^{24}&=&-({\dot v}+2K{\dot u})s^{14}+2{\dot x}s^{34},\label{eq22}.
\end{eqnarray}
From Eq. (\ref{eq11}) we obtain
\begin{eqnarray}
\frac{d^2u}{d\tau^2}&=&0,\label{eq23}\\
\frac{d^2x}{d\tau^2}&=&\frac{1}{2m}K_{xx}{\dot u}s^{13}-
\frac{1}{2}K_x{\dot u}^2,\label{eq24}\\
\frac{d^2y}{d\tau^2}&=&\frac{1}{2m}K_{yy}{\dot u}s^{14}-
\frac{1}{2}K_y{\dot u}^2,\label{eq25}.
\end{eqnarray}
The other component of the trajectory may be found by considering $H=-\frac{m}{2}$
from which, Eq. (\ref{eq4}) results in
\begin{equation}
{\dot v}=\frac{1}{\dot u}(1-K+{\dot x}^2+{\dot y}^2),\label{eq26}
\end{equation}
We fix the gauge such that the solution to Eq. (\ref{eq23}) takes the following form
\begin{equation}
u=\tau.\label{eq27}
\end{equation}
Also Eqs. (\ref{eq17})-(\ref{eq18}) result respectively in
\begin{eqnarray}
s^{13}(\tau)&=&{\mbox constant}=\alpha,\label{eq28}\\
s^{14}(\tau)&=&{\mbox constant}=\beta.\label{eq29}
\end{eqnarray}
So Eqs. (\ref{eq24})-(\ref{eq25}) become
\begin{eqnarray}
\ddot x&=&\frac{\alpha}{m}h\omega^2\sin(\omega\tau)-h\omega^2\sin(\omega\tau)x
-h\omega^2\cos(\omega\tau)y,\label{eq30}\\
\ddot y&=&-\frac{\beta}{m}h\omega^2\sin(\omega\tau)-h\omega^2\cos(\omega\tau)x
+h\omega^2\cos(\omega\tau)y,\label{eq31}
\end{eqnarray}
respectively. With $\zeta=x+iy$ and $c=\alpha-i\beta$ the solution to the above
system of coupled differential equations is
\begin{eqnarray}
\zeta(\tau)&=&\frac{\bar c}{2m}+\frac{2ihc}{m(4-h^2)}e^{i\omega\tau}+
\frac{{\bar c}h^2}{2m(4-h^2)}e^{-2i\omega\tau}+pe^{ia\omega\tau}\nonumber\\
&+&\frac{ih}{(1+a)^2}{\bar p}e^{-i(1+a)\omega\tau}+qe^{ib\omega\tau}
+\frac{ih}{(1+b)^2}{\bar q}e^{-i(1+b)\omega\tau}
,\label{eq32}
\end{eqnarray}
where $a=\frac{-1+\sqrt{1+4h}}{2}$, $b=\frac{-1+\sqrt{1-4h}}{2}$
with $p,q$ constants determined by initial conditions, and overbar
denoting complex conjugation. For $h\leq\frac{1}{4}$ both $x$ and
$y$ components contain $\sin$ and $\cos$ terms only, resulting in
a combined oscillation in the $x-y$ plane. For $h>\frac{1}{4}$
exponential terms appear in both $x$ and $y$. In this case
oscillatory terms are modulated by either exponentially decreasing
or increasing terms and the resulting motion differs significantly
from the previous type. We do not investigate this type of motion
here.

It is also interesting to find expressions for the Grassmannian variables $\psi^\mu$.
Using Eqs. (\ref{eq2}), (\ref{eq10}), and (\ref{eq15}) we obtain
\begin{eqnarray}
\psi^1(\tau)&=&\mbox{a constant Grassmannian}\label{eq1f}\\
\psi^2(\tau)&=&\psi^1(-2K-\frac{\dot v}{\dot u}+\frac{2\dot x}{\dot u}f(\tau)
+\frac{2\dot y}{\dot u}g(\tau))\label{eq2f}\\
\psi^3(\tau)&=&\psi^1f(\tau)\label{eq3f}\\
\psi^4(\tau)&=&\psi^1g(\tau)\label{eq4f}
\end{eqnarray}
where $f(\tau)$ and $g(\tau)$ are solutions to $df(\tau)/d\tau=-1/2K_x\dot u$
and $dg(\tau)/d\tau=-1/2K_y\dot u$ respectively. The immediate result of these
expressions is the vanishing of the dual supercharge and the chiral charge:
$$Q^\star=\Gamma_\star=0$$
Having the exact solution to Eqs. (\ref{eq17})-(\ref{eq18}) and
(\ref{eq23})-(\ref{eq25}), it is now straightforward to find
solutions to Eqs. (\ref{eq19})-(\ref{eq22}) and (\ref{eq26}). So
we have found in principal the exact solutions to the equations
for the trajectory and the spin of the particle. The resulting
expressions are algebraically involved. To get physical insight
into the nature of the resulting solutions, one may use a weak
field limit of the solutions. We have gathered in the appendix
some of the linearized solutions.
\section{The world-line deviations}
From (\ref {eq15}) and (\ref{eqd2}) we obtain
\begin{eqnarray*}
\frac{dJ^{13}}{d\tau}=\frac{dJ^{14}}{d\tau}=0
\end{eqnarray*}
whose solutions are
\begin{eqnarray}
J^{13}(\tau)&=&\mbox{constant}=\epsilon\label{eqd3},\\
J^{14}(\tau)&=&\mbox{constant}=\delta\label{eqd4}.
\end{eqnarray}
Equations (\ref{eqd1}) results in
\begin{equation}\label{eqd5}
\frac{d^2n^1}{d\tau^2}=0
\end{equation}
One may choose the following solution to the above equation
\begin{equation}
n^1(\tau)=0
\end{equation}
Now inserting the above relations in  equations (\ref{eqd1}) we
get
\begin{eqnarray}
\frac{d^2n^3(\tau)}{d\tau^2}&=&+h\omega^2\sin(\omega\tau)n^3(\tau)+
\frac{\epsilon}{m}h\omega^2\sin(\omega\tau),\label{eqd6}\\
\frac{d^2n^4(\tau)}{d\tau^2}&=&-h\omega^2\sin(\omega\tau)n^4(\tau)-
\frac{\delta}{m}h\omega^2\sin(\omega\tau).\label{eqd7}
\end{eqnarray}
The other component, $n^2(\tau)$ may be obtained from the
condition $\dot{x}\cdot \frac{Dn}{D\tau}=0$. An approximate
solution to (\ref{eqd6}) and (\ref{eqd7}) is
\begin{eqnarray}
n^3(\tau)&=&A+h((\frac{\epsilon}{m}-A)(\omega\tau-\sin(\omega\tau))
),\label{eqd8}\\
n^4(\tau)&=&B+h((-\frac{\delta}{m}+B)(\omega\tau-\sin(\omega\tau))
),\label{eqd9}
\end{eqnarray}
in which $A,B$ denote the initial separations of the two spinning
particles at rest near the origin of the coordinates with the same
$\frac{\epsilon}{m}$ and $\frac{\delta}{m}$. These show how the
relative accelerations of spinning particles depend on the spin.
The terms independent of the spin are due to the usual tidal
force.

\section{Discussion}
\begin{figure}[h]
\centering \epsffile{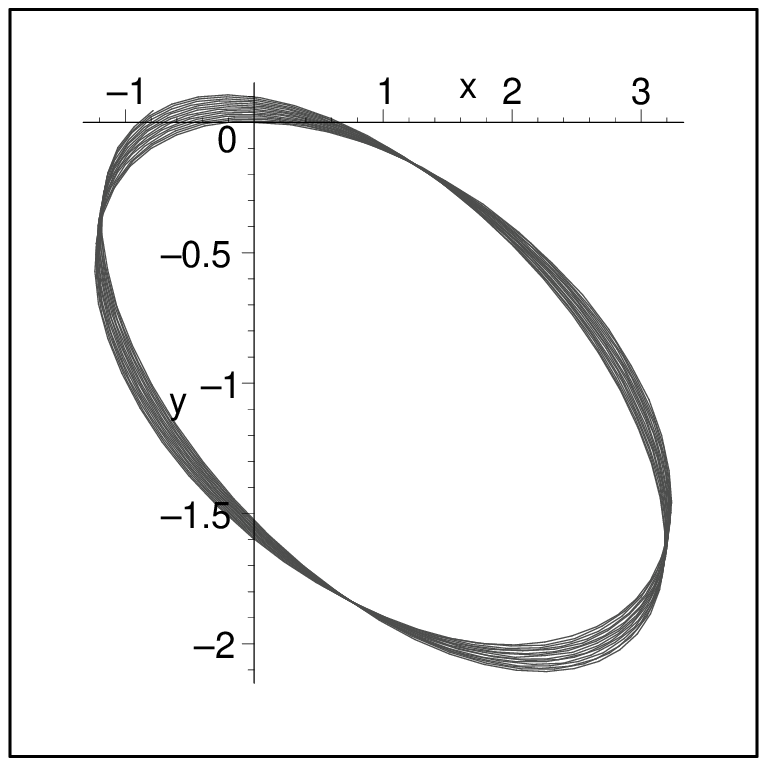} \caption{} The particle's
trajectory projected onto transverse plane for
$\frac{\alpha}{m}=\frac{\beta}{m}=2,\omega=1,h=0.001$ and
$0\leq\tau\leq 100000$. \label{fig1}
\end{figure}

The projection of the particle's trajectory onto the $x-y$ plane
is given by (\ref{eq32}). This is depicted for a particle
initially at rest at the origin in Fig.\ref{fig1} for some set of
the parameters. As the figure shows the path is an ellipse with a
slow precession. The dimensions of this ellipse depend on the
particles's spin. For a given frequency, this precession gets
slower for smaller values of $h$.

Using a weak field approximation (see the appendix) give some
insight into the nature of the exact solution we found. Let us
examine the trajectory and spin of the particle in some rather
special situations. For a particle at rest at the origin of
coordinates at $\tau=0$ (which we consider as the time in which
the wave arrives), if the initial orientation of the spin is
chosen to be $\alpha=\beta=0$, and $\gamma\neq 0$ then
\begin{equation}
(u(\tau),v(\tau),x(\tau),y(\tau))=(\tau,\tau,0,0),\label{eq41}
\end{equation}
that is the particle remains at rest at the initial coordinates, and
\begin{equation}
(s^1(\tau),s^2(\tau),s^3(\tau),s^4(\tau))=(-\gamma,\gamma,0,0),
\label{eq42}
\end{equation}
its spin remains unchanged. These are in agreement with the results of \cite{1}
and \cite{7}.

For $\gamma=0$ and $\alpha\neq 0$, $\beta\neq 0$, the particle's trajectory is
\begin{eqnarray}
x(\tau)&=&\frac{\alpha}{2m}-\frac{h\alpha}{m}\sin(\omega\tau)
+\frac{\alpha}{m}\sin(h\omega\tau)-\frac{\alpha}{2m}\cos(h\omega\tau)
,\label{eq43}\\
y(\tau)&=&\frac{\beta}{2m}+\frac{h\beta}{m}\sin(\omega\tau)
-\frac{\beta}{m}\sin(h\omega\tau)-\frac{\beta}{2m}\cos(h\omega\tau)
,\label{eq44}\\
v(\tau)&=&\tau.\label{eq45}
\end{eqnarray}
and its spin is
\begin{eqnarray}
s^{1}(\tau)&=&\frac{2h\omega}{m}\alpha\beta(1-\cos(\omega\tau)),\label{eq46}\\
s^2(\tau)&=&\frac{2h\omega}{m}\alpha\beta(1-\cos(\omega\tau)),\label{eq47}\\
s^3(\tau)&=&\beta,\label{eq48}\\
s^4(\tau)&=&-\alpha.\label{eq49}
\end{eqnarray}
It is now clear from Eqs. (\ref{eq43})-(\ref{eq45}) that the
particle's displacement along the direction of the wave
propagation ($z=\frac{v-u}{2}$) is much smaller ($O(h^2)$) than
its displacement in the transverse plane, even though this
transverse displacement is practically very small ($O(h)$) itself.
This is similar to the behaviour of spinless particles.

For general initial spin orientation ($\alpha\neq 0, \beta\neq 0,\gamma\neq 0$),
the transverse components of the particle's spin have oscillations of very
small amplitudes (of the order of
$h$) around their initial orientation. The component of the spin
parallel to the direction of the wave propagation $s_z=\frac{s^2-s^1}{2}$
remains unaffected, again in agreement with \cite{1}.

The relative  acceleration of two nearby spinning particles is
given by (\ref{eqd8}) and (\ref{eqd9}). They show that the
relative acceleration oscillates with the frequency of the wave.

As argued in \cite{2} we may recover Eqs. (\ref{eq9}) and
(\ref{eq11}) -(\ref{eq12}) from the so called Papapetrou-Dixon
equations by neglecting the terms quadratic in spin. On the other
hand, as we saw in our calculations above, the contribution from
spin are usually small. Thus it seems neglecting the very small
contributions from terms quadratic in spin makes no significant
loss of physical content. Hence in the lack of exact solutions to
Papapetrou-Dixon equations in gravitational wave spacetime, our
results may be generalized to them as well (at least in spacetime
regions far from the strong sources of these waves).

\clearpage \noindent{\bf acknowledgment}

I would like to thank the research council of the Payame Noor
University for partial financial support.
\section*{appendix}
In this appendix we gather some linearized solutions for the
particles trajectory and spin. Putting
$\zeta(0)=\dot{\zeta}(0)=0$, Eq. (\ref{eq32}) reduces to first
order in $h$ to
\begin{eqnarray}
\zeta(\tau)&=&\frac{\bar c}{2m}+\frac{ihc}{2m}e^{i\omega\tau}-
\frac{ihc}{2m}e^{-i\omega\tau}-\frac{{\bar
c}+2ic}{4m}e^{ih\omega\tau} +\frac{2ic-\bar
c}{4m}e^{-ih\omega\tau} ,\label{eq32ff}
\end{eqnarray}
or equivalently
\begin{eqnarray}
x(\tau)&=&\frac{\alpha}{2m}-\frac{h\alpha}{m}\sin(\omega\tau)
+\frac{\alpha}{m}\sin(h\omega\tau)-\frac{\alpha}{2m}\cos(h\omega\tau)
,\label{eq33}\\
y(\tau)&=&\frac{\beta}{2m}+\frac{h\beta}{m}\sin(\omega\tau)
-\frac{\beta}{m}\sin(h\omega\tau)-\frac{\beta}{2m}\cos(h\omega\tau)
.\label{eq34}
\end{eqnarray}
Thus from Eq. (\ref{eq26}) we obtain (with $v(0)=0$)
\begin{eqnarray}
v(\tau)&=&\tau+O(h^2).\label{eq35}
\end{eqnarray}
Eq. (\ref{eq27}) and (\ref{eq33})-(\ref{eq35}) determine the
particle's trajectory. From Eq. (\ref{eq19}) we obtain
\begin{eqnarray}
s^{34}(\tau)&=&\gamma \label{eq36}
\end{eqnarray}
where $\gamma$ is a constant of integration.

We may now put Eqs. (\ref{eq28})-(\ref{eq29}) and (\ref{eq36})
into (\ref{eq20})-(\ref{eq22}) to find the other components of
$s^{\mu\nu}$. The results may be fed into (\ref{eq13}) (with
$\epsilon^{1234}=+1$) to give $s^\mu$
\begin{eqnarray}
s^{1}(\tau)&=&\frac{2h\omega}{m}\alpha\beta(1-\cos(\omega\tau))-\gamma,
\label{eq37}\\
s^2(\tau)&=& \frac{2h\omega}{m}\alpha\beta(1-\cos(\omega\tau))+\gamma,\label{eq38}\\
s^3(\tau)&=&\beta-\frac{h\omega}{m}\alpha\gamma(1-\cos(\omega\tau))
,\label{eq39}\\
s^4(\tau)&=&-\alpha-\frac{h\omega}{m}\beta\gamma(1-\cos(\omega\tau))
,\label{eq40}.
\end{eqnarray}

\end{document}